\documentclass[reprint,aps,prd,onecolumn,notitlepage,showpacs,nofootinbib,preprintnumbers,superscriptaddress]{revtex4-1}
\usepackage{amsmath}
\usepackage{amsfonts}
\usepackage{amssymb}
\usepackage{latexsym}
\usepackage{color}
\usepackage{graphicx}
\usepackage{bm}
\usepackage{epsfig}
\usepackage[english]{babel}
\usepackage{hyperref}
\usepackage{times}
\usepackage{comment}

\def\diag{\mbox{diag}}

\begin{document}

\title{Fractional relativity}
\author{Tower Wang}
\email[Electronic address: ]{twang@phy.ecnu.edu.cn}
\affiliation{Department of Physics, East China Normal University,\\
Shanghai 200241, China\\ \vspace{0.2cm}}
\date{\today\\ \vspace{1cm}}
\begin{abstract}
By fractional relativity we mean a theoretical framework to study physics with the dispersion relation $E^{\alpha}=m^{\alpha}c^{2\alpha}+p^{\alpha}c^{\alpha}$, which recovers special relativity at $\alpha=2$. One such framework is established in a particular curved energy-momentum space. It is shown that the fractional Schr\"{o}dinger equation arises as a nonrelativistic limit of the Klein-Gordon equation in fractional relativity. In this framework, the relative locality makes no contribution to the position uncertainty at the classical level, and the Faraday's law in classical electrodynamics is modified by fractional derivatives.
\end{abstract}


\maketitle




\section{Introduction}\label{sect-intro}
It is well known that the standard Schr\"{o}dinger equation in quantum mechanics can be derived from the Feynman path integral over the Brownian paths. In a series of pioneering works \cite{2000PhLA..268..298L,2000PhRvE..62.3135L,2002PhRvE..66e6108L}, by studying the path integral over the L\'{e}vy paths, Laskin formulated a generalization of standard quantum mechanics, dubbed the fractional quantum mechanics. In this theory, the basic equation is the so-called fractional Schr\"{o}dinger equation
\begin{equation}\label{fSE}
i\hbar\frac{\partial\psi(\mathbf{r},t)}{\partial t}=D_{\alpha}\left(-\hbar^2\triangle\right)^{\alpha/2}\psi(\mathbf{r},t)+V(\mathbf{r},t)\psi(\mathbf{r},t),
\end{equation}
where $\left(-\hbar^2\triangle\right)^{\alpha/2}$ is the quantum Riesz fractional derivative, $1<\alpha\leq2$, and $[D_{\alpha}]=\mathrm{erg}^{1-\alpha}\times\mathrm{cm}^{\alpha}\times\mathrm{sec}^{-\alpha}$ in CGS units. In the special case at $\alpha=2$, $D_2=1/(2m)$, it recovers the standard quantum mechanics. In the literature, fractional quantum mechanics has been studied from various perspectives. Some of them \cite{2004JMP....45.3339N,2007JMP....48d3502W,2011arXiv1103.3295B,2011PhLA..375.3541L,2017CSF...102...16L} are devoted to generalizations of equation \eqref{fSE} with a fractional time derivative, but none has tried the limit of special relativity. In this paper, we are interested in a covariant generalization of fractional quantum mechanics, namely the fractional relativistic quantum mechanics.

The Hamiltonian of equation \eqref{fSE} implies the kinetic energy \cite{2002PhRvE..66e6108L}
\begin{equation}\label{T}
T=D_{\alpha}p^{\alpha}
\end{equation}
with the momentum $p=|\mathbf{p}|=(\mathbf{p}\cdot\mathbf{p})^{1/2}$. We observe that the following form of dispersion relation
\begin{equation}\label{disp}
E^{\alpha}=m^{\alpha}c^{2\alpha}+p^{\alpha}c^{\alpha},
\end{equation}
has the nonrelativistic limit
\begin{equation}
E\simeq mc^2+\frac{mc^2}{\alpha}\left(\frac{p}{mc}\right)^{\alpha},
\end{equation}
which reproduces equation \eqref{T} if one identifies
\begin{equation}\label{Dm}
D_{\alpha}=\frac{mc^2}{\alpha m^{\alpha}c^{\alpha}}.
\end{equation}
Note that the dispersion relation \eqref{disp} has the massless limit $E=pc$, which will be interpreted as the dispersion relation of photon in this paper, then the speed of light remains to be $c$. It is easy to show that $v=dE/dp\leq c$ for $\alpha>1$.

Inspired by this observation, we are also interested in a suitable framework for studying physical consequences of the elegant relation \eqref{disp}. As the title shows, such a framework is named fractional relativity. When fractional relativity deviates from special relativity, its nonrelativistic limit also deviates from Newtonian mechanics as indicated by equation \eqref{T}, and the corresponding classical electrodynamics deviates from Maxwell's theory.

In section \ref{sect-fKG}, we will study a fractional relativistic Lagrangian for scalar field and take the nonrelativistic limit to get the fractional Schr\"{o}dinger equation. In section \ref{sect-geom}, we will interpret the fractional relativistic Lagrangian from the viewpoint of curved energy-momentum space \cite{2011PhRvD..84h4010A,2013arXiv1307.7988A}. In such a space, a massless vector field obeys the deformed Maxwell equations, which will be presented in section \ref{sect-fED}. The paper will be concluded with a brief discussion in section \ref{sect-disc}. From now on, we will always assume equation \eqref{Dm} and work in units in which $\hbar=c=1$.

\section{Fractional Schr\"{o}dinger equation from fractional relativity}\label{sect-fKG}
In accordance with dispersion relation \eqref{disp}, the Lagrangian density for a scalar field $\phi(\mathbf{x},t)$ of mass $m$ has the form
\begin{equation}\label{lagdr}
\mathcal{L}=\frac{1}{2m^{\alpha-2}}\dot{\phi}(-\partial_{t}^2)^{\alpha/2-1}\dot{\phi}-\frac{1}{2m^{\alpha-2}}\nabla\phi\cdot(-\triangle)^{\alpha/2-1}\nabla\phi-\frac{1}{2}m^2\phi^2.
\end{equation}
Here the Riesz fractional derivatives with respect to time and 3-dimensional space are defined respectively as
\begin{eqnarray}\label{Riesz}
\nonumber&&(-\partial_{t}^2)^{\alpha/2}\phi(\mathbf{x},t)=\frac{1}{2\pi}\int dEe^{-iEt}E^{\alpha}\int dt'e^{iEt'}\phi(\mathbf{x},t'),\\
&&(-\triangle)^{\alpha/2}\phi(\mathbf{x},t)=\frac{1}{(2\pi)^3}\int d\mathbf{p}e^{i\mathbf{p}\cdot\mathbf{x}}|\mathbf{p}|^{\alpha}\int d\mathbf{x}'e^{-i\mathbf{p}\cdot\mathbf{x}'}\phi(\mathbf{x}',t).
\end{eqnarray}
From these definitions, one can prove the formulas for integration by parts \cite{2002PhRvE..66e6108L} and subsequently a deformed Klein-Gordon equation
\begin{equation}\label{dKG}
\frac{1}{m^{\alpha-2}}\left[(-\partial_{t}^2)^{\alpha/2}-(-\triangle)^{\alpha/2}\right]\phi=m^2\phi.
\end{equation}

In the nonrelativistic limit, the kinetic energy $T$ is much smaller than the rest mass $m$. Therefore, it makes sense to rewrite
\begin{equation}\label{phi-psi}
\phi(\mathbf{x},t)=\frac{1}{\sqrt{2m}}\left[e^{-imt}\psi(\mathbf{x},t)+e^{imt}\psi^{\ast}(\mathbf{x},t)\right]
\end{equation}
with $\psi$ oscillating much more slowly than $e^{-imt}$ in time. If we use $\simeq$ to denote equivalence up to terms suppressed by $T/m$ and terms with a rapidly oscillating phase factor $e^{\pm2imt}$ \cite{2015PhRvD..92j3513G,2016PhRvL.117l1801B}, then we can prove
\begin{eqnarray}
\label{lagdr3}&&\phi^2\simeq\frac{1}{m}\psi^{\ast}\psi,\\
\label{lagdr1}&&\dot{\phi}(-\partial_{t}^2)^{\alpha/2-1}\dot{\phi}\simeq m^{\alpha-1}\psi^{\ast}\psi+m^{\alpha-3}\dot{\psi}\dot{\psi}^{\ast}+im^{\alpha-2}\left(\dot{\psi}\psi^{\ast}-\psi\dot{\psi}^{\ast}\right),\\
\label{lagdr2}&&\nabla\phi\cdot(-\triangle)^{\alpha/2-1}\nabla\phi\simeq\frac{1}{2m}\left[\nabla\psi^{\ast}\cdot(-\triangle)^{\alpha/2-1}\nabla\psi+\nabla\psi\cdot(-\triangle)^{\alpha/2-1}\nabla\psi^{\ast}\right].
\end{eqnarray}
Plugging them into equation \eqref{lagdr}, we obtain the effective Lagrangian density in the nonrelativistic limit
\begin{equation}\label{lagdnr}
\mathcal{L}\simeq\frac{i}{2}\left(\dot{\psi}\psi^{\ast}-\psi\dot{\psi}^{\ast}\right)-\frac{1}{4m^{\alpha-1}}\left[\nabla\psi^{\ast}\cdot(-\triangle)^{\alpha/2-1}\nabla\psi+\nabla\psi\cdot(-\triangle)^{\alpha/2-1}\nabla\psi^{\ast}\right].
\end{equation}
We have dropped the second term in equation \eqref{lagdr1} because it is small compared to the last term. Utilizing the formula for integration by parts \cite{2002PhRvE..66e6108L}, it is straightforward to variate this Lagrangian to get the fractional Schr\"{o}dinger equation \eqref{fSE} with $V=0$.

The Lagrangian density \eqref{lagdr} describes a free particle. It can be augmented with a self-interaction term and an external potential
\begin{equation}
\mathcal{L}=\frac{1}{2m^{\alpha-2}}\dot{\phi}(-\partial_{t}^2)^{\alpha/2-1}\dot{\phi}-\frac{1}{2m^{\alpha-2}}\nabla\phi\cdot(-\triangle)^{\alpha/2-1}\nabla\phi-\frac{1}{2}m^2\phi^2-\frac{\lambda}{4}\phi^4-U(\mathbf{x},t)\phi^2.
\end{equation}
Up to total derivative terms, its nonrelativistic limit is
\begin{equation}
\mathcal{L}\simeq\frac{i}{2}\left(\dot{\psi}\psi^{\ast}-\psi\dot{\psi}^{\ast}\right)-\frac{1}{2m^{\alpha-1}}\nabla\psi^{\ast}\cdot(-\triangle)^{\alpha/2-1}\nabla\psi-\frac{\lambda}{4m^2}\left(\psi^{\ast}\psi\right)^2-\frac{U}{m}\psi^{\ast}\psi.
\end{equation}
The fractional Schr\"{o}dinger equation \eqref{fSE} can be derived from this Lagrangian by setting $\lambda=0$, $U=mV$.

\section{Geometric interpretation}\label{sect-geom}
It is unlikely for the fractional relativity to settle in general relativity with the Riemannian metric. It seems natural but turns out also difficult to explain the dispersion relation \eqref{disp} with the Finsler metric. Instead, we have succeeded in accommodating it in a curved energy-momentum space. The idea of curved energy-momentum space is invented by Amelino-Camelia et al in reference \cite{2011PhRvD..84h4010A} and explored later in reference \cite{2013arXiv1307.7988A}. It is enlightening to recast the Lagrangian density \eqref{lagdr} in an apparently covariant form
\begin{equation}\label{lagdcov}
\mathcal{L}=\frac{1}{2}(\partial_{\mu}\phi)g^{\mu\nu}\partial_{\nu}\phi-\frac{1}{2}m^2\phi^2,
\end{equation}
where the inverse metric is
\begin{equation}\label{inmetric}
g^{\mu\nu}=\frac{1}{m^{\alpha-2}}\diag\left[(-\partial_{t}^2)^{\alpha/2-1},-(-\triangle)^{\alpha/2-1},-(-\triangle)^{\alpha/2-1},-(-\triangle)^{\alpha/2-1}\right].
\end{equation}
One way to understand the dispersion relation \eqref{disp} is taking equation \eqref{inmetric} as a metric of momentum space \cite{2011PhRvD..84h4010A}
\begin{equation}\label{pcmetric}
dm^2=g^{\mu\nu}dp_{\mu}dp_{\nu}=\frac{1}{\left(E^{\alpha}-|\mathbf{p}|^{\alpha}\right)^{(\alpha-2)/\alpha}}\left(E^{\alpha-2}dE^2-|\mathbf{p}|^{\alpha-2}|d\mathbf{p}|^2\right),
\end{equation}
in which $|d\mathbf{p}|=(d\mathbf{p}\cdot d\mathbf{p})^{1/2}$. In the spherical coordinates,
\begin{equation}\label{psmetric}
dm^2=\frac{1}{\left(E^{\alpha}-p^{\alpha}\right)^{(\alpha-2)/\alpha}}\left(E^{\alpha-2}dE^2-p^{\alpha-2}dp^2-p^{\alpha}d\Omega^2\right)
\end{equation}
where $d\Omega^2=d\vartheta^2+\sin^2\vartheta d\varphi^2$ is the metric on a unit 2-sphere. Then the dispersion relation \eqref{disp} can be established if we interpret the particle mass as the geodesic distance from the origin \cite{2011PhRvD..84h4010A}, $m=D(p,0)$. This will be done as follows.

Akin to the geodesics in coordinate space, the metric geodesics in momentum space obey the geodesic equation \cite{2013arXiv1307.7988A}
\begin{equation}
\frac{d^2p_{\lambda}}{d\varepsilon^2}+\Gamma^{\mu\nu}_{\lambda}\frac{dp_{\mu}}{d\varepsilon}\frac{dp_{\nu}}{d\varepsilon}=0,
\end{equation}
where $\Gamma^{\mu\nu}_{\lambda}$ is the Levi-Civita connection, and $\varepsilon$ is the affine parameter. Restricted to metric \eqref{psmetric}, in the simplest case, it is enough for us to focus on radial geodesics, i.e. $d\Omega^2=0$. Then the geodesic equation takes the form
\begin{eqnarray}
\nonumber\frac{E''}{E}-\frac{(\alpha-2)p^{\alpha}}{2\left(E^{\alpha}-p^{\alpha}\right)}\left(\frac{E'}{E}-\frac{p'}{p}\right)^2=0,\\
\frac{p''}{p}+\frac{(\alpha-2)E^{\alpha}}{2\left(E^{\alpha}-p^{\alpha}\right)}\left(\frac{E'}{E}-\frac{p'}{p}\right)^2=0.
\end{eqnarray}
Here primes denote derivatives with respect to $\varepsilon$. Solving these equations, we find a radial geodesic
\begin{equation}
E=C_0\varepsilon,\quad p=C_1\varepsilon,\quad\vartheta=C_2,\quad\varphi=C_3.
\end{equation}
Inserting it into the line element \eqref{psmetric}, we work out the geodesic distance from the origin
\begin{equation}
\int_{0}^{\varepsilon}d\varepsilon\sqrt{g^{\mu\nu}\frac{dp_{\mu}}{d\varepsilon}\frac{dp_{\nu}}{d\varepsilon}}=\left(C_0^{\alpha}-C_1^{\alpha}\right)^{2/\alpha}\varepsilon.
\end{equation}
The left-hand side is mass $m$ by definition \cite{2011PhRvD..84h4010A}, while the right-hand side is equal to $\left(E^{\alpha}-p^{\alpha}\right)^{2/\alpha}$. This establishes the dispersion relation \eqref{disp}.

In terms of metric \eqref{inmetric}, the deformed Klein-Gordon equation \eqref{dKG} is equivalent to
\begin{equation}\label{cKG}
-\frac{1}{\sqrt{-g}}\partial_{\mu}\left[\sqrt{-g}g^{\mu\nu}(\partial_{\nu}\phi)\right]=m^2\phi.
\end{equation}
It coincides in form with the covariant Klein-Gordon equation in general relativity. However, here $g^{\mu\nu}$ is the metric in energy-momentum space. One can regard this equation as the covariant Klein-Gordon equation in curved energy-momentum space. As an immediate consequence, the fractional Schr\"{o}dinger equation can be considered as the nonrelativistic limit of the Klein-Gordon equation in curved energy-momentum space.

In reference \cite{2011PhRvD..84h4010A}, it was discovered that in curved energy-momentum space, different observers see different spacetimes, and the spacetimes they observe are dependent of energy and momentum. The coordinates of particles involved in an interaction removed from the origin of the observer by a vector $z^{\lambda}$ are spread over a region of order $\Delta x^{\mu}\sim z^{\lambda}\Gamma^{\mu\nu}_{\lambda}p_{\nu}$. For the metric \eqref{pcmetric}, we find $\Gamma^{\mu\nu}_{\lambda}p_{\nu}=0$ generally. As a result, although the energy-momentum space is curved here, it does not introduce any nonlocality $\Delta x^{\mu}$ at the classical level. Of course, at the quantum level, there will still be a nonlocality proportional to the Planck constant \cite{2016PhRvE..93f6104L}.

\section{Classical electrodynamics in fractional relativity}\label{sect-fED}
In the above, we have employed equations \eqref{inmetric} and \eqref{pcmetric} interchangeably. However, interpreted as the geodesic distance from the origin \cite{2011PhRvD..84h4010A}, the mass $m$ should not appear in the metric directly. Therefore, equation \eqref{pcmetric} is more healthy than equation \eqref{inmetric}, especially in the relativistic limit. For massive particles, the deformed Klein-Gordon equation \eqref{dKG}, corresponding to \eqref{inmetric}, is equivalent to the covariant Klein-Gordon equation \eqref{cKG}. But for massless particles, it should be superseded by the fractional Klein-Gordon equation
\begin{equation}
-\left[(-\partial_{t}^2)^{\alpha/2}-(-\triangle)^{\alpha/2}\right]^{2/\alpha}\phi=0.
\end{equation}
This equation is obtained by substituting metric \eqref{pcmetric} into equation \eqref{cKG} and setting $m=0$ on the right-hand side.

The lesson is useful for studying electrodynamics in the same momentum space, if we take photons as a massless vector field according to dispersion relation \eqref{disp}. Supposedly the Maxwell's equations in curved energy-momentum space is formally the same as in general relativity,
\begin{equation}\label{cMax}
\frac{1}{\sqrt{-g}}\partial_{\lambda}\left[\sqrt{-g}g^{\lambda\mu}g^{\kappa\nu}(\partial_{\mu}A_{\nu}-\partial_{\nu}A_{\mu})\right]=4\pi J^{\kappa}.
\end{equation}
Recalling that $A^{\lambda}=(\Phi,\mathbf{A})$ and $J^{\lambda}=(\rho,\mathbf{J})$, we can see the scalar potential $\Phi$ and the vector potential $\mathbf{A}$ satisfy
\begin{eqnarray}\label{dMax}
\nonumber-\frac{(-\triangle)^{\alpha/2}}{\left[(-\partial_{t}^2)^{\alpha/2}-(-\triangle)^{\alpha/2}\right]^{(\alpha-2)/\alpha}}\Phi+\frac{(-\partial_{t}^2)^{(\alpha-2)/2}}{\left[(-\partial_{t}^2)^{\alpha/2}-(-\triangle)^{\alpha/2}\right]^{(\alpha-2)/\alpha}}\partial_{t}(\nabla\cdot\mathbf{A})&=&-4\pi\rho,\\
\left[(-\partial_{t}^2)^{\alpha/2}-(-\triangle)^{\alpha/2}\right]^{2/\alpha}\mathbf{A}-\frac{(-\triangle)^{(\alpha-2)/2}}{\left[(-\partial_{t}^2)^{\alpha/2}-(-\triangle)^{\alpha/2}\right]^{(\alpha-2)/\alpha}}\nabla(\nabla\cdot\mathbf{A}+\partial_{t}\Phi)&=&-4\pi\mathbf{J}.
\end{eqnarray}

As will be clarified soon, equation \eqref{dMax} is invariant under certain gauge transformation on $\Phi$ and $\mathbf{A}$. The quantities of physical significance should be gauge-invariant. In classical electrodynamics, physically significant quantities are the electric field $\mathbf{E}$ and the magnetic field $\mathbf{B}$. Their six components are the elements of the field-strength tensor $F^{\lambda\kappa}=g^{\lambda\mu}g^{\kappa\nu}(\partial_{\mu}A_{\nu}-\partial_{\nu}A_{\mu})$. In the matrix form, that is
\begin{equation}
F^{\lambda\kappa}=\left(\begin{array}{cccc}
0&-E_x&-E_y&-E_z\\
E_x&0&-B_z&B_y\\
E_y&B_z&0&-B_x\\
E_z&-B_y&B_x&0\\
\end{array}\right).
\end{equation}
Because $\partial_{\lambda}(\sqrt{-g})=0$, equation \eqref{cKG} is simply $\partial_{\lambda}F^{\lambda\kappa}=4\pi J^{\kappa}$, which immediately recovers the Coulomb's law $\nabla\cdot\mathbf{E}=4\pi\rho$
as well as the Amp\`{e}re's law with a displacement current $\nabla\times\mathbf{B}-\partial_{t}\mathbf{E}=4\pi\mathbf{J}$.
Comparing them with equation \eqref{dMax}, we infer that
\begin{eqnarray}
\label{EA}\mathbf{E}&=&-\frac{(-\triangle)^{(\alpha-2)/2}}{\left[(-\partial_{t}^2)^{\alpha/2}-(-\triangle)^{\alpha/2}\right]^{(\alpha-2)/\alpha}}\nabla\Phi-\frac{(-\partial_{t}^2)^{(\alpha-2)/2}}{\left[(-\partial_{t}^2)^{\alpha/2}-(-\triangle)^{\alpha/2}\right]^{(\alpha-2)/\alpha}}\partial_{t}\mathbf{A},\\
\label{BA}\mathbf{B}&=&\frac{(-\triangle)^{(\alpha-2)/2}}{\left[(-\partial_{t}^2)^{\alpha/2}-(-\triangle)^{\alpha/2}\right]^{(\alpha-2)/\alpha}}\nabla\times\mathbf{A}.
\end{eqnarray}
The divergence of equation \eqref{BA} implies the absence of free magnetic poles, $\nabla\cdot\mathbf{B}=0$.
The curl of the equation \eqref{EA} yields
\begin{equation}\label{Faraday}
\nabla\times\mathbf{E}+\frac{(-\partial_{t}^2)^{(\alpha-2)/2}}{(-\triangle)^{(\alpha-2)/2}}\partial_{t}\mathbf{B}=0.
\end{equation}
This is similar to the familiar Faraday's law, and slightly different by a fractional derivative operator.

Despite of the complicated form of equation \eqref{dMax}, we see most of the Maxwell's equations for $\mathbf{E}$ and $\mathbf{B}$ take the same form as they are in standard classical electrodynamics, except for that the Faraday's law gets modified. In a vacuum, $\rho=0$ and $\mathbf{J}=0$, these equations can be combined to give the propagation equation of electromagnetic waves
\begin{eqnarray}
\nonumber\left[(-\partial_{t}^2)^{\alpha/2}-(-\triangle)^{\alpha/2}\right]\mathbf{E}&=&0,\\
\left[(-\partial_{t}^2)^{\alpha/2}-(-\triangle)^{\alpha/2}\right]\mathbf{B}&=&0.
\end{eqnarray}

It is easy to check that the electric field \eqref{EA} and the magnetic field \eqref{BA} are invariant under the following ``gauge'' transformation
\begin{eqnarray}
\nonumber\mathbf{A}&\rightarrow&\mathbf{A}+(-\triangle)^{(\alpha-2)/2}\nabla\Lambda,\\
\Phi&\rightarrow&\Phi-(-\partial_{t}^2)^{(\alpha-2)/2}\partial_{t}\Lambda.
\end{eqnarray}
One may also confirm that equation \eqref{dMax} is invariant under this transformation. Therefore, equation \eqref{dMax} can be simplified with one more condition, such as the Lorenz gauge condition $\nabla\cdot\mathbf{A}+\partial_{t}\Phi=0$ or the Coulomb gauge condition $\nabla\cdot\mathbf{A}=0$.

\section{Discussion}\label{sect-disc}
Both fractional quantum mechanics and curved energy-momentum space are unconventional interesting ideas, but they grew up in very different fields. This paper puts them together in the fractional relativity with dispersion relation $E^{\alpha}=m^{\alpha}c^{2\alpha}+p^{\alpha}c^{\alpha}$. We hope the future development in curved energy-momentum space will shed light on a better understanding of fractional quantum mechanics and fractional classical electrodynamics. Conversely, more numerical and experimental work on fractional quantum mechanics \cite{2015PhRvL.115r0403Z} and fractional classical electrodynamics is imperative to simulate the curved energy-momentum space, just like analogue models of gravity \cite{2005LRR.....8...12B}. We leave it as an open problem to embed the fractional relativity into a theoretical framework other than curved energy-momentum space.

\begin{acknowledgments}
This work is supported partially by the National Natural Science Foundation of China (Grant No. 91536218). The author would like to thank Rui-Yan Yu for enlightening discussions.
\end{acknowledgments}

\end{document}